\documentstyle[12pt]{article}
\begin{document}
\thispagestyle{empty}
\def\cqkern#1#2#3{\copy255 \kern-#1\wd255 \vrule height #2\ht255
depth  #3\ht255 \kern#1\wd255}
\def\cqchoice#1#2#3#4{\mathchoice%
   {\setbox255\hbox{$\rm\displaystyle #1$}\cqkern{#2}{#3}{#4}}%
{\setbox255\hbox{$\rm\textstyle #1$}\cqkern{#2}{#3}{#4}}%
{\setbox255\hbox{$\rm\scriptstyle #1$}\cqkern{#2}{#3}{#4}}%
   {\setbox255\hbox{$\rm\scriptscriptstyle #1$}\cqkern{#2}{#3}{#4}}}
\def\CC{\mathord{\cqchoice{C}{0.65}{0.95}{-0.1}}}
\def\x{\stackrel{\otimes}{,}}
\def\y{\stackrel{\circ}{\scriptstyle\circ}}
\def\proof{\noindent Proof. \hfill \break}
\def\a{\begin{eqnarray}}
\def\b{\end{eqnarray}}
\def\o{\overline}
\def\p{{1\over{2\pi i}}}
\def\Q{{\scriptstyle Q}}
\def\P{{\scriptstyle P}}
\def\t{\textstyle}
\begin{flushright}
ENSLAPP/L/433/93
\\

 \end{flushright}
\centerline{\LARGE Superhamiltonian formalism}
\centerline{\LARGE for $2D$ $N=1,2$ theories}
\vspace{1truecm}
\vskip0.5cm
\centerline{{\large E. Ivanov}}
\vskip.5cm
\centerline{Bogoliubov Theoretical Laboratory, JINR,}
\centerline{Dubna, Head Post Office, P.O. Box 79,}
\centerline{101000, Moscow, Russia}
\vskip1cm
\centerline{{\large F. Toppan}}
\vskip.5cm
\centerline{Laboratoire de Physique Th\'{e}orique ENSLAPP,}
\centerline{Ecole
Normale Sup\'erieure de Lyon,}
\centerline{46 All\'ee d'Italie, 69007 Lyon, France}
\vskip.5cm
\vskip1.5cm
\centerline{\bf Abstract}
We show how to formulate $2$-dimensional
supersymmetric $N=1,2$
theories, both massive and conformal, within a manifestly
supersymmetric hamiltonian framework, via the introduction of
a (super)-Poisson brackets structure defined on superfields.
In this approach, as distinct from the previously
known superfield hamiltonian formulations,
the dynamics is not separated into two unrelated $2D$ light-cone
superspaces, but is recovered by specifying boundary conditions at
a  given ``super-time" coordinate. So the approach proposed
provides a natural generalization
of canonical hamiltonian formalism. One of its interesting features
is
that the physical and auxiliary fields equations appear on equal
footing
as the Hamilton ones.
\vskip.5cm
\newpage\setcounter{page}1
\section{Introduction}
The most convenient and efficient way to deal with
supersymmetric theories is a manifestly supersymmetric
approach based on
the concept of superfields. This fact has been recognized since the
early
works on supersymmetry, and manifestly supersymmetric superfield
lagrangian
formalisms are of common use (see \cite{GGRS} and references
therein).
On the other hand, the hamiltonian methods of treating field
theories,
though being advantageous in a number of aspects (e.g., in what
concerns
the quantization), received little attention
in the framework of the superfield approach.

A version of the superfield hamiltonian formalism for supersymmetric
$2D$ theories is already available (see, e.g., \cite{Isaev},
\cite{ChaichianLukierski}).
However, it has been formulated for the two
separated $1$-dimensional dynamics in the two unrelated
$2D$ light-cone superspaces and so seems to be of immediate use
mainly in superconformally-invariant integrable theories which admit
(sometimes after a field redefinition) on-shell separation into
independent
left and right movers. Examples of that kind are supplied by super
Liouville and super Toda theories.

In this paper we introduce another superfield Poisson brackets
structure which directly generalizes the canonical (not the
light-cone)
bosonic Poisson structure and is applicable not only to
superconformally-invariant $N=1$ and $N=2$, $2D$ models, but also
to their massive deformations, such as the super sine-Gordon and
sinh-Gordon models. We show that the dynamics of
$N=1,\; 2$ supersymmetric $2D$ theories can be readily recovered
within
the superhamiltonian framework based on such a super-Poisson
brackets
structure and that it takes a form very similar to the dynamics of
bosonic
$2D$ theories in the standard hamiltonian formulation.

The dynamics of our theory is recovered in terms of boundary
conditions
which are naturally expressed by specifying the values of
the superfields at a given ``supertime" coordinate, instead of
specifying their values on two separated light-cone supersurfaces.

Our framework seems therefore the natural set-up for studying
properties of integrable models like the exchange algebra (this was
indeed the original motivation for our work
\cite{{Toppan},{TopZhang}}).

Though we could formulate our $2D$ super-Poisson brackets formalism
in a complete generality, we will specialize here to
the three simplest non-trivial integrable field-theoretical examples,
namely,
the $N=1$ (massive) super sinh-Gordon theory \cite{Babelon},
the $N=1$ superconformal
affine Liouville theory (super-CAL) \cite{TopZhang}, and finally
the $N=2$ super sine-Gordon theory \cite{Evans}.

We decided to treat here the case of supersymmetric $N=1,2$
sin-Gordon theories instead of the super Liouville ones in order to
stress
the fact that our formalism can be applied to massive models as
well.
The super-CAL theory has been chosen as an example of superconformal
theory
which involves in particular a pair of superfields having momenta
conjugate to each other.

The generalization of this formulation to more complicated theories
is
completely
straightforward.

\section{$N=1$ superhamiltonian framework}
Our starting point in this Section will be the algebra of $N=1$, $2D$
supersymmetry. Its nonvanishing structure relations in the standard
light-cone notation and in the realization via the $N=1$, $2D$
superspace
coordinates $(z^{++}, z^{--}, \theta^{+}, \theta^{-})$ read
\begin{equation}
(Q_{+})^2 = P_{++}\;, \;\;\; (Q_{-})^2 = P_{--}\;.
\end{equation}
Here
$$
Q_{\pm} = i {\partial\over {\partial \theta^{\pm}}} +
\theta^{\pm}
\partial_{\pm \pm} \;, \;\;\; P_{\pm \pm} = i\partial_{\pm \pm}\;.
$$
We will also need the algebra of $N=1$ spinor covariant derivatives
\begin{eqnarray}
D_{\pm} &=& {\partial\over {\partial \theta^{\pm}}} +
i \theta^{\pm}\partial_{\pm \pm}\;,
\label{CD} \\
(D_+)^{2} \;=\; i\partial_{++}\;, && (D_{-})^2 \;=\;
i\partial_{--}\;,
\;\; \{ D_+, D_- \} = 0\;. \label{ACD}
\end{eqnarray}

As the first step in developing a hamiltonian formalism one should
choose which of the $2D$ coordinates will be identified
with the evolution parameter, the time. One possibility is to take as
such a
coordinate $z^{++}$ or
$z^{--}$; these options amount to the light-cone hamiltonian
formalism.
Two other possibilities lead to the canonical $2D$ hamiltonian
approach
a superfield extension of which we are going to construct.
They correspond to the following two alternative identifications of
the
time-like and spatial $2D$ coordinates
\begin{eqnarray} \label{A}
x &=& {\textstyle {1\over 2}}(z^{++}+z^{--})
\nonumber\\
t &=& {\textstyle {1\over 2}}(z^{++}-z^{--})
\end{eqnarray}
or
\begin{eqnarray} \label{B}
t &=& {\textstyle {1\over 2}}(z^{++}+z^{--})
\nonumber\\
x &=& {\textstyle {1\over 2}}(z^{++}-z^{--})\;.
\end{eqnarray}
In what follows we will keep to the former definition as it has been
used in ref. \cite{TopZhang} which was just a first incitement for
undertaking the present investigation. We will comment on
the second, more customary option
in the end of this section.

Along with introducing the coordinates $t$ and $x$ as in eq.
(\ref{A}),
we also pass to the new ``rotated'' fermionic coordinates
\a
\theta^1&=& {\textstyle {1\over 2}}(\theta^++\theta^-)\nonumber\\
\theta^0&=& {\textstyle {1\over 2}}(\theta^+ -\theta^-)\;.
\b
Defining the ``rotated'' spinor derivatives by
\a
&&D_0=D_+-D_-\quad\quad D_1 = D_+ +D_-,\nonumber
\b
\a
D_0 &=& {\partial\over\partial\theta^0} +i
( \theta^1\partial_t+\theta^0\partial_x)\nonumber\\
D_1 &=& {\partial\over\partial\theta^1} +i
( \theta^1\partial_x+\theta^0\partial_t)\;,
\b
one can rewrite the algebra (\ref{ACD}) in the following equivalent
form
\a
&& {D_0}^2 ={D_1}^2 = i\partial_x\;, \quad\quad \{D_0,D_1\} = 2i
\partial_t
\label{}
\b
Note the useful relation
\a
&& D_+D_-={\textstyle {i\over 2}}\partial_t - {\textstyle {1\over
2}}D_1D_0\;.
\label{Rel}
\b
We also present the transformation laws of the coordinates
$t,x,\theta^1,\theta^0$ under $N=1$ supersymmetry:
\a \label{STr}
\delta \theta^1 = \epsilon^1\;, \; \delta \theta^0 = \epsilon^0\;, \;
\delta x = -i(\epsilon^1\theta^1 +\epsilon^0\theta^0)\;,\;
\delta t = -i(\epsilon^1\theta^0 + \epsilon^0\theta^1)\;.
\b

Now we possess all the necessary matter to approach our task of
constructing a superfield version of the hamiltonian formalism. As
a sample model we take the super sinh-Gordon theory which is
introduced
via the action
\a \label{action}
{\cal S} &\equiv& \int d^2 z d^2 \theta L = \int d^2 z d^2{\theta}
\{ D_+\Phi {D_-\Phi}+i\alpha e^{\Phi}+i\beta e^{-\Phi} \}\;,
\b
where $\Phi$ is a bosonic superfield. Without loss of generality
the real parameter $\alpha$ can be set equal to unity
through a shift of the superfield $\Phi$, while $\beta$ is a mass
parameter
which measures the breaking of conformal invariance. The super
Liouville
theory is recovered by letting $\beta\rightarrow 0$.

The equation of motion is given by
\a  \label{eqmo}
D_+D_-\Phi&=& {\textstyle{i\over 2}}(e^{\Phi}-\beta e^{-\Phi})\;,
\b
or, with making use of (\ref{Rel}),
\a \label{eqmo1}
\dot{\Phi} &=& - i D_1 D_0 \Phi + (e^{\Phi}-\beta e^{-\Phi})\;.
\b
By applying the derivative $D_0$ on both sides of the latter
equation,
one gets the important consequence
\a \label{eqmo2}
D_0\dot{\Phi} &=& D_1\Phi ' - D_0\Phi (e^{\Phi}+\beta e^{-\Phi})\;.
\b
Here we denoted,
as usual, ${\dot\Phi} \equiv\partial_t\Phi,\; \Phi'\equiv
\partial_x\Phi $.

Now we wish to show that eqs.(\ref{eqmo1}), (\ref{eqmo2}) can be
re-derived
as the first-order Hamilton equations in the framework of the
appropriate
superhamiltonian formalism.

To this end, let us first rewrite the action (\ref{action}) so as to
explicitly single out $\dot{\Phi}$. This can be achieved by taking
one
of spinor derivatives, say $D_1$, off the integration measure in
${\cal S}$ and throwing it on the integrand\footnote{We define
Berezin integrals as
$$ \int d^2 z d^2\theta \equiv \int d^2 x D_+D_- = -{1\over 2}
\int d^2 z D_1 D_0\;, \int d^2 z d \theta^1 \equiv \int d^2 z D_1\;.
$$}
\a
{\cal S}&=& \int d^2 z d{\theta^1} {\cal L} \label{Snew} \\
{\cal L} &=& {\textstyle {1\over 2}} i{\dot{\Phi}} D_0\Phi
-{\textstyle {1\over 4}}i\Phi ' D_1\Phi -
{\textstyle {1\over 4}}D_0\Phi D_1D_0 \Phi - {\textstyle {i\over 2}}
D_0\Phi ( e^{\Phi}-\beta e^{-\Phi})\;. \label{Lag}
\b
All the $\theta^0$- dependent terms in the lagrangian ${\cal L}$
are reduced to $x, t$-derivatives and so do not contribute to the
integral
in (\ref{Snew}), though the involved superfields can still be
regarded as
given on the whole $N=1$ superspace $(t, x, \theta^0, \theta^1)$.
In other
words, without loss of anything we may treat the superfield argument
$\theta^0$ in ${\cal L}$ as some cyclic Grassmann variable which can
be
fixed at any ``value'' we wish (in particular, it can be put equal to
zero).
Note that in the action (\ref{Snew}) there remains only one manifest
supersymmetry
\begin{equation} \label{MS}
\delta \theta^1 = \epsilon^1\;, \;\;\; \delta x = -i\epsilon^1
\theta^1
\end{equation}
(the time coordinate can be made inert under
$\epsilon^1$-supertranslations
by the redefinition $t\rightarrow
\hat{t} = t +i\theta^1\theta^0$). Supertranslations with the
parameter
$\epsilon^0$ now cannot be realized as pure coordinate
transformations;
they mix $\Phi$ with $D_0\Phi$ and $\dot{\Phi}$.

Next natural step is to define, in the standard way, the superfield
momentum $\Pi_{\Phi}$ canonically conjugate to $\Phi$
\a
\Pi_{\Phi}&=& {\textstyle{\delta {\cal S} \over \delta {\dot{\Phi}}
}} =
{\textstyle{\partial {\cal L} \over \partial {\dot{\Phi}} }} =
{\textstyle {i\over 2}} D_0\Phi\;.
\b
Surprisingly, the conjugate momentum turns out to be fermionic. This
is of course a consequence of the fact that the superfield lagrangian
${\cal L}$ is a fermionic object in the present case.

Now, again following the text-book prescriptions, we introduce the
superhamiltonian
\a
H &=& \int dxd\theta^1\{ \Pi_{\Phi}\cdot {\dot{\Phi}} -
{\cal L}\} =\nonumber\\
&=&\int dxd\theta^1 \{ {\textstyle {1\over 4}}i\Phi ' D_1 \Phi
- \Pi_\Phi D_1\Pi_\Phi + \Pi_\Phi
(e^{\Phi}-\beta e^{-\Phi})\}\;. \label{SH}
\b
Integration in (\ref{SH}) over the superplane $\{x,\theta^1\}$ is
quite
natural because the latter is the minimal superextension of the
line $\{ x \}$ closed under $x$ translations and ``$N=1/2$''
supersymmetry (\ref{MS}).

Using the equation of motion (\ref{eqmo1}) and
its corollary (\ref{eqmo2}) one easily proves the conservation laws
\a  \label{CL}
D_0 H &=&\dot{H} =0\;,
\b
which state that {\it on shell} $H$ does not depend on the
coordinates
$\theta^0$ and $t$. Then it is natural to join these coordinates into
a
``supertime'' $T \equiv (t, \theta^0)$ and, respectively, to identify
the
remaining set of coordinates as the ``superspatial'' coordinate
$X \equiv (x, \theta^1)$. Thus, instead of the Poisson brackets at
equal
time appearing on the component level, in the superfield hamiltonian
formalism we are led to define the Poisson brackets at equal
supertime.
Note that the superhamiltonian density (the integrand in (\ref{SH}))
includes explicit $\theta^0$ and the derivative $\partial_t$ which
enter
via the spinor derivative $D_1$. In order to avoid their appearance
and so to
be able to integrate by part with respect to $D_1$ one should
redefine
$t$ as follows
\a
t \Rightarrow \tilde{t} = t -i\theta^1\theta^0\;.
\b
In this basis
$$
D_1 = {\partial \over {\partial \theta^1}} + i\theta^1 \partial_x\;.
$$

In accord with the last remark we define the super-Poisson brackets
at equal
supertime $\tilde{T} \equiv
(\tilde{t}, \theta^0)$.
The only non-vanishing bracket is
\a \label{SP}
\{\Pi_{\Phi} (X, \tilde{T}), \Phi (X', \tilde{T}') \}_{\tilde{T}
\equiv \tilde{T}'}
&=& \Delta (X, X') \equiv
\delta (x-x')(\theta^1 -{\theta^1}' )\;.
\label{SPoisson}
\b
Here $\Delta (X, X')$ is the supersymmetric delta function on the
superplane $\{ X \}$:
$$
\int d^2 X \Delta (X, Y) F(X) = F(Y)\;,\;\; \;d^2X \equiv dx
d\theta^1\;,
$$
with $F(X)$ being an arbitrary ``$N=1/2$'' superfield. Note that the
equality
of supertimes, $\tilde{T}=\tilde{T}'$, amounts to the following
relations
between the original coordinates
\a  \label{Sim}
\theta^0 = {\theta^0}'\;, \;\;\;t-t' -i (\theta^1 - {\theta^1}')\;
\theta^0 = 0\;.
\b
These relations are invariant both under the $\epsilon^0$ and
$\epsilon^1$
supertranslations (\ref{STr}), which justifies our choice of
``super-simultaneity'' in (\ref{SP}).

Now it is an easy exercise to check that the Hamilton equations
pertinent
to the superhamiltonian (\ref{SH}) and the super-Poisson structure
(\ref{SP})
\a
{\dot{\Phi}} &=&  \{ H, \Phi\}\nonumber\\
{\dot \Pi}_\Phi &=&  \{ H, \Pi_\Phi \}
\b
just reproduce the equations of motion (\ref{eqmo1}), (\ref{eqmo2}).

The hamiltonian (\ref{SH}) satisfying the super-conservation laws
(\ref{CL}) and the super-Poisson structure (\ref{SP}) are
the basic ingredients of the superhamiltonian formalism for $N=1$,
$2D$
theories. It is straightforward to check that after performing the
integration over $d\theta^1$ and elimination of the auxiliary field
in
(\ref{SH}) the latter becomes the standard component hamiltonian,
while (\ref{SP}) gives rise to the conventional Poisson brackets for
the
physical component fields present in the $\theta$ expansions of
$\Phi$, $\Pi$.
These basic structures are not specific just for the $N=1$
sine-Gordon model:
they can be equally defined for any other $N=1$, $2D$ theory and
shown to
possess similar properties.

To demonstrate this, let us apply the previous construction to
the $N=1$ superconformal affine theory of ref. \cite{TopZhang}.

The action
$$
{\cal S} = \int d^2 z d^2 \theta L  \equiv \int d^2 z d \theta^1
{\cal L}
$$
is now given by
\a
{\cal S} = \int d^2 z d^2{\theta}
\{{\textstyle {1\over 2}}D_0\Phi D_1\Phi +
{\textstyle {1\over 2}}D_0\Sigma D_1\Lambda +
{\textstyle {1\over 2}} D_0\Lambda D_1 \Sigma +
i\gamma (e^{\Phi} + e^{\Lambda-\Phi})\}\;,
\b
with $\Phi, \Sigma , \Lambda $ being independent real $N=1$
superfields.
The equations of motion are
\a   \label{eqmo4}
{\dot \Phi}&=& -i D_1D_0 \Phi + \gamma ( e^{\Phi}-e^{\Lambda -\Phi} )
\nonumber \\
{\dot \Lambda} &=& - i D_1D_0 \Lambda \nonumber\\
{\dot\Sigma } &=& -i D_1D_0 \Sigma + \gamma e^{\Lambda -\Phi}
\b
As before, we integrate in ${\cal S}$ over $d\theta^0$ and define the
conjugate momenta as follows
\a
\Pi_{\Phi}&=& {\textstyle{\delta {\cal S} \over \delta {\dot{\Phi}}
}}
= {\textstyle{\partial {\cal L} \over \partial {\dot{\Phi}} }}
= {\textstyle {i\over 2}}
D_0\Phi\;, \nonumber \\
\Pi_{\Sigma}&=& {\textstyle{\delta {\cal S} \over \delta
{\dot{\Sigma}} }}
= {\textstyle{\partial {\cal L} \over \partial {\dot{\Sigma}} }} =
{\textstyle {i\over 2}}
D_0\Lambda\;, \nonumber\\
\Pi_{\Lambda}&=& {\textstyle{\delta {\cal S} \over \delta
{\dot{\Lambda}} }}
= {\textstyle{\partial {\cal L} \over \partial {\dot{\Lambda}} }} =
{\textstyle {i\over 2}}
D_0\Sigma\;.
\b
Denoting with $\Psi$ a generic superfield, the super-Poisson brackets
are
introduced, at equal supertime $\tilde{T}$, through
\a
\{\Pi_{\Psi} (X, T), \Psi (X',T') \}_{\tilde{T} \equiv \tilde{T}'}
&=&
\Delta (X, X')
\b
The superhamiltonian in this case is given by
\a
H &=& \int dxd\theta^1\{ \Pi_{\Phi}\cdot {\dot{\Phi}}
+
\Pi_{\Sigma}\cdot {\dot{\Sigma}}
+
 \Pi_{\Lambda}\cdot {\dot{\Lambda}} - {\cal L}\} \nonumber\\
&=&\int dxd\theta^1 \{ {\textstyle {1\over 4}}i\Phi ' D_1 \Phi
 + {\textstyle {1\over 4}}i\Sigma ' D_1 \Lambda
 + {\textstyle {1\over 4}}i\Lambda ' D_1 \Sigma
+ \gamma \Pi_{\Phi}e^{\Phi} + \gamma (\Pi_{\Sigma}-\Pi_{\Phi} )
e^{\Lambda -\Phi}\nonumber\\
&&
-\Pi_\Phi D_1\Pi_\Phi - \Pi_\Sigma D_1 \Pi_\Lambda -
\Pi_\Lambda D_1\Pi_{\Sigma} \}\;.
\b
The equations of motion are reproduced by the Hamilton system
\a
{\dot{\Psi}} &=&  \{ H, \Psi\}\nonumber\\
{\dot \Pi}_{\Psi} &=& \{ H, {\Pi}_{\Psi}\}\;.
\b
It is easy to check that the conservation laws (\ref{CL}) are valid
in this case too.

Let us comment now on the alternative definition of $t$ and $x$ as in
eq.(\ref{B}). For this choice one cannot construct an analogous
superhamiltonian formulation, though the component one in terms
of physical fields certainly exists. The reason is clear from the
anticommutation relations (8) (or the relations between the
supertranslation generators which are basically the same).
Under the choice we kept to until now there exists a $N=1/2$
subalgebra
$(Q_1, P_x)$ in $N=1$ SUSY algebra, extending the $x$ translation
generator $P_x$. Just due to this fact one can define a ``spatial''
superplane
$(\theta_1, x)$ which is closed under this $N=1/2$ SUSY and plays
a crucial role in
the above construction. The $N=1/2$ SUSY is the only manifest
supersymmetry of our superhamiltonian formalism.
On the other hand,
the choice of $t,x$ as in eq. (5) amounts to the interchange of
$\partial_t$ and $\partial_x$ in eq. (8) and one can easily see
that in this case there is no any kind of $N=1/2$ subalgebra
extending $P_x$. Therefore one cannot single out a closed subspace in
$N=1$, $2D$ superspace, so as it would involve only $x$ in its
bosonic sector.

However, the choice (5) is distinguished by the positivity
arguments: indeed, in this case
$$
P_t = (Q_+)^2 + (Q_-)^2
$$
and the energy is strictly positive as a consequence of
the $N=1$, $2D$ SUSY algebra. There is no such a remarkable
property for the choice (4). Nevertheless, it is an easy effort
to see that both choices are physically equivalent. Let us
start with the component off-shell form of the action (11),
assuming that the choice (4) has been done and the
superfield $\Phi$ has the following $\theta$ expansion
\a \label{Exp}
\Phi(x,t,\theta^+, \theta^-) = \phi(x,t) + \theta^+ \psi_+(x,t) +
\theta^- \psi_-(x,t) +{1\over 2} \theta^+\theta^- F(x,t)\;.
\b
Then one can check that the formal substitutions
\a \label{Subst}
\psi_- \Rightarrow i\psi_-,\; \; F \Rightarrow iF,\;\;
\alpha \Rightarrow i\alpha,\;\; \beta \Rightarrow i\beta
\b
take this component action just into the form pertinent to the choice
(5) and
vice versa. Thus we can use our superhamiltonian formalism at all
stages of
computations and at the final stage, in order to ensure the
positiveness of
energy, accomplish the substitutions (\ref{Subst}). Note that in
terms
of superfields the replacements (\ref{Subst}) correspond to invoking
another type of reality
condition for $\Phi$
$$
\Phi^{\dagger} (z, \theta^+, \theta^-) = \Phi (z, \theta^+,
-\theta^-)\;,
$$
which means that $\Phi$ is assumed to be real with respect to the
new involution defined as the product of ordinary complex conjugation
and
reflection of $\theta_-$. Note that the latter reflection is an
obvious automorphism of the $N=1$ superalgebra (1). Just owing to
this property, the alternative definition of reality is still
compatible with
the $N=1$, $2D$ supersymmetry.

Let us conclude this section by mentioning an application of our
super- Poisson
bracket formalism: it can be used to analyze the
integrability property of the $N=1$
super sinh-Gordon and superCAL theory: let a supersymmetric Lax pair
(see ref.\cite{TopZhang})
be defined through the positions
\begin{equation}
{ L}_+=D_+\Psi+e^{{\rm ad}\Psi}{\cal E}_+,~~~~{
L}_-=-D_-\Psi+
e^{-{\rm ad}\Psi}{\cal E}_-
\end{equation}
where  $\Psi$ is a superfield taking value in the Cartan subalgebra
of $osp(2|2)^{(2)}$ (respectively ${\hat {osp(2|2)^{(2)}}}$)
and ${\cal E}_+$, ${\cal E}_-$ are the sum of odd positive and
negative simple root vectors of the same superalgebra. The
zero-curvature condition for the above Lax pair is equivalent to the
equations of motion
derived in the superhamiltonian framework. The integrability
property
of such theories are made manifest by the appearance of the classical
$r$-matrices $r^\pm$ (they are defined in terms of the superalgebra
generators and
their explicit expression is given in \cite{TopZhang})
through the relation for
${ L}={ L}_+ +{ L}_-$:
\a
\relax \{{ L}_1 (X,T), {L}_2 (Y,T)\}&=& [
r^{\pm} , { L}_1 (X,T) +{ L}_2 (Y,T) ] \delta (X,Y)
\label{rmatrix}
\b
where ${ L}_1\equiv { L}\otimes 1, { L}_2\equiv 1\otimes
{ L}$. The l.h.s. is referred to the super-Poisson bracket.
Notice
in particular the presence of the (super-)conjugate momentum
$ \Pi_\Psi \equiv (D_+-D_-)\Psi $.
In the super-Poisson bracket formalism the above equality
(\ref{rmatrix})
is directly proven, with no need of performing lenghty computation at
the level of component fields.

\section{The $N=2$ superhamiltonian framework}

In this section we extend the construction of the super-Poisson
brackets
and superhamiltonian formalism to the $N=2$ case, taking as an
illustration the $N=2$ super sine-Gordon theory.

The two-dimensional $N=2$ superspace is parametrized by
$z^{\pm\pm}$ and the complex
fermionic coordinates
$\theta^\pm$ (and their conjugate ${\overline{\theta}^\pm}$). We will
sometimes use the notation $Z = (z^{++}, z^{--}, \theta^+,
{\overline{\theta}^+}, \theta^-, {\overline{\theta}^-})$. The
$N=1$
superspace is recovered by letting
${{\theta^\pm}}={\overline{\theta}^\pm}$.
The $N=2$ spinor derivatives $D_{\pm},{\overline D}_{\pm}$ are
defined as:
\a
D_{\pm} &=&{\partial\over{\partial\theta^{\pm}}}-
i{\overline \theta}^{\pm}\partial_{\pm\pm}\nonumber\\
{\overline D}_{\pm}
&=&-{\partial\over\partial{\overline\theta}^{\pm}}+
i{ \theta}^{\pm}\partial_{\pm\pm} \;.
\b
The only non-vanishing bracket between them is given by
\a
\{D_\pm, {\overline D}_\pm \} &=& 2i \partial_{\pm\pm} \;.\nonumber
\b
In particular we have
\a
{D_\pm}^2 = {{\overline D}_{\pm}}^2 =0&& \;.\nonumber
\b
Just as in the previous section, it is convenient to introduce
the ``rotated" spinor derivatives $D_{0,1}$, ${\overline D}_{0,1}$
defined as
\a
D_0&=&D_+-D_-\nonumber\\
&=& {\partial\over\partial\theta^0}-i{\o\theta}^1\partial_t -
i{\o \theta}^0\partial_x\nonumber\\
D_1&=&D_++D_-\nonumber\\
&=& {\partial\over\partial\theta^1}-i{\o\theta}^0\partial_t -
i{\o \theta}^1\partial_x\nonumber\\
{\o D}_0&=&-
 {\partial\over\partial{\o\theta}^0}+i{\theta}^1\partial_t +
i{ \theta}^0\partial_x\nonumber\\
{\o D}_1&=&
- {\partial\over\partial{\o\theta}^1}+i{\theta}^0\partial_t +
i{ \theta}^1\partial_x\;.
\b
They satisfy the following anticommutation relations
\a  \label{ACD22}
\{D_0,{\o D}_0 \}\;=\;\{D_1,{\o D}_1\} &=& 2i\partial_x\nonumber\\
\{D_0,{\o D}_1\} \;=\; \{D_1, {\o D}_0\} &=& 2i\partial_t\nonumber\\
\{D_{i}, D_{j}\} \;=\; \{{\o D}_{i}, {\o D}_{j}\} &=& 0 \;\;\;\;\;
(i,j = 0,1).
\b
An $N=2$ chiral superfield $\Phi$ is introduced by the constraint
\a  \label{CC}
{{D}_\pm}\Phi = 0, &&
\b
while its conjugate satisfies
\a  \label{ACC}
{\overline D}_\pm \Phi^{\dagger} =0
\b
and so it is an anti-chiral $N=2$ superfield:
\a
\Phi \equiv \Phi (z^{++}_R, z^{--}_R, {\o \theta}^{+}, {\o
\theta}^{-})\;,
\;\; \Phi^{\dagger} \;\equiv \; \Phi^{\dagger} (z^{++}_L, z^{--}_L,
{\theta}^{+},
{\theta}^{-})\;,
\b
\a
z^{\pm\pm}_R = z^{\pm\pm} + i\theta^{\pm}{\o \theta}^{\pm}\;, \;\;
z^{\pm\pm}_L \;= \; z^{\pm\pm} - i\theta^{\pm}{\o \theta}^{\pm}\;.
\b
In what follows, the parametrizations of
$N=2$ superspace using the bosonic coordinates $z^{\pm \pm}_L$ and
$z^{\pm \pm}_R$ will be referred to, respectively, as the $Z_L$ and
$Z_R$ ones.

The action ${\cal S}$ of the super sinh-Gordon theory reads
\cite{Evans}:
\a  \label{S2}
{\cal S} &=& \int d^2 z d^4\theta \{\Phi \Phi^\dagger\} +
\int d^2 z_R d^2{\o\theta}
\{e^\Phi +\beta e^{-\Phi} \} + \int d^2 z_L d^2\theta
\{e^{\Phi^\dagger}+
\beta e^{-\Phi^\dagger}\}
\b
where we defined the $N=2$ superspace integration measures by
\a  \label{Int2}
\int d^2 z d^4\theta &=& \int d^2 z_L d^2\theta {\o D}_0 {\o D}_1 =
\int d^2 z_R d^2{\o\theta}D_1D_0 \nonumber \\
\int d^2 z_L d^2\theta &=& \int d^2 z_L D_1D_0\;, \;\;
\int d^2 z_R d^2{\o \theta} \;= \; \int d^2 z_R {\o D}_0{\o D}_1\;.
\b
For what follows it will be convenient to rewrite the action as
\a \label{SChL}
{\cal S}
&=& S_L + S_R \equiv \int d^2 z_L d^2 \theta L_L +
\int d^2 z_R d^2 {\o \theta}_R L_R \nonumber \\
&=& {\textstyle {1\over 2}}\int d^2 z_L d^2 \theta \{ {\o D}_0{\o
D}_1\Phi
\cdot \Phi^\dagger + 2(e^{\Phi^\dagger} +\beta e^{-\Phi^\dagger}
)\nonumber\\
&& +{\textstyle {1\over 2}}\int d^2z_R d^2{\o\theta} \{
\Phi\cdot D_1D_0\Phi^\dagger
+ 2(e^{\Phi} +\beta e^{-\Phi} )\}\;.
\b

The equations of motion following from (\ref{S2}) are
\a \label{eqmoN2}
D_1D_0\Phi^\dagger &=& -e^\Phi +\beta e^{-\Phi}\nonumber\\
{\o D}_0{\o D}_1 \Phi &=& -e^{\Phi^\dagger} + \beta e^{-\Phi^\dagger}
\b
Acting on them by spinor derivatives one gets the important
consequences
\a
D_0{\Phi^{\dagger}}'-D_1{\dot\Phi}^{\dagger} &=& {\textstyle{i\over
2}}\;
{\o D}_1\Phi \; (e^{\Phi}+\beta e^{-\Phi})
\nonumber\\
D_0{\dot \Phi}^{\dagger} - D_1{\Phi^{\dagger}}' &=&
{\textstyle{i\over 2}}\; {\o D}_0\Phi \;
(e^{\Phi}+\beta e^{-\Phi})\label{Con1} \\
{\o D}_1\dot{\Phi} -{\o D}_0 \Phi ' &=& {\textstyle{i\over 2}}\;
D_1\Phi^{\dagger}\; (e^{\Phi^\dagger}+\beta e^{-\Phi^\dagger})
\nonumber\\
{\o D}_1\Phi ' -{\o D}_0\dot{\Phi} &=&
{\textstyle{i\over 2}}\; D_0\Phi^{\dagger} \;
(e^{\Phi^{\dagger}}+\beta e^{-\Phi^\dagger}) \label{Con2}
\b
Note that in the present case, as it follows from the anticommutation
relations (\ref{ACD22}), one cannot cast the equations of motion
(\ref{eqmoN2})
in the form analogous to (\ref{eqmo1}). However, using (\ref{ACD22})
and
the chirality conditions (\ref{CC}) and (\ref{ACC}), one finds the
algebraic identities
\a \label{Al}
\dot{\Phi} = -{i\over 2} D_1 {\o D}_0 \Phi \;, \;\;\;
\dot{\Phi}^{\dagger} = -{i\over 2} {\o D}_1 D_0 \Phi^{\dagger}\;,
\b
which in the $N=2$ superhamiltonian formalism replace the dynamical
relation (\ref{eqmo1}).

After integration in both chiral parts of the action over,
respectively,
$d \theta^0$, $d {\o\theta}^0$ one gets
\a
{\cal S} &\equiv& \int d^2 z_L d\theta^1{\cal L}_L + \int d^2 z_R
d{\o \theta}^1 {\cal L}_R \nonumber \\
&=& \int d^2z_Ld\theta^1 \{ i ({\o D}_1 \Phi)'\Phi^{\dagger} +
i({\o D}_0\Phi){\dot\Phi}^{\dagger}
-{\textstyle {1\over 2}} {\o D}_1 ({\o D}_0\Phi )(D_0\Phi^{\dagger})+
D_0\Phi^{\dagger} (e^{\Phi^{\dagger}}-\beta e^{-\Phi^\dagger})
\}\nonumber
\\
&-&\int d^2z_Rd{\o\theta}^1 \{ i (D_1\Phi^{\dagger})'\Phi +i
(D_0\Phi^{\dagger})
{\dot\Phi}
-{\textstyle{1\over 2}} D_1(D_0\Phi^{\dagger})({\o D}_0\Phi )-
{\o D}_0 \Phi (e^{\Phi}-\beta e^{-\Phi})\}.\nonumber\\
&&
\b
The super-conjugate momenta are defined to be
\a  \label{SM2}
\Pi_{\Phi^{\dagger}} &=& {\textstyle {\delta
{\cal S}\over \delta {\dot \Phi}^{\dagger}}} = 2i {\o D}_0
\Phi\nonumber\\
\Pi_{\Phi} &=& {\textstyle {\delta {\cal S}\over \delta {\dot
\Phi}}}=
-2iD_0 \Phi^{\dagger} = (\Pi_{\Phi^{\dagger}})^{\dagger}\;.
\b
Notice that, when computing, e.g., $\Pi_{\Phi^\dagger}$, one should
rewrite the kinetic part of ${\cal S}$ in the basis
where ${\dot \Phi}^\dagger$ is
unconstrained ($z_L,\theta^1$ basis). Then both parts of the action
give the same contribution, which explains the factor $2$ in
(\ref{SM2}).

The $N=2$ super-Poisson brackets are defined at equal left-chiral or
right-chiral
supertimes $T_L \equiv (t_L, \theta^0)$, $T_R \equiv (t_R, {\o
\theta}^0)$.
With $X_L \equiv (x_L, \theta^1)$, $X_R \equiv (x_R, {\o \theta}^1)$,
the only non-vanishing brackets are
\a \label{PB2}
\{\Pi_{\Phi} (Z_R), \Phi ({Z'}_R ) \}_{T_R \equiv {T_R}'} &=&
\Delta (X_R, {X'}_R)\nonumber\\
\{ \Pi_{\Phi^{\dagger}} (Z_L),
\Phi^{\dagger}({Z'}_L)\}_{T_L \equiv {T_L}'} &=&  {\o \Delta }
(X_L, {X'}_L)
\b
where $\Delta (X_R, Y_R)$, ${\o \Delta }(X_L, Y_L)$ are the chiral
delta-functions on the conjugate ``superspatial'' planes
$x_R, {\o \theta}^1$ and $x_L, \theta^1$. Note that both sides of
(\ref{PB2}) identically vanish under $D_0\; ({\o D}_0), D_1\; ({\o
D}_1)$
acting on the second superargument and under $D_0\;({\o D}_0)$ acting
on the first one. This follows from the chirality of $\Phi$ and the
definition
of conjugate momenta. At the same time, $D_1 \; ({\o D}_1)$,
while acting on the first superargument in the l.h.s. of (\ref{PB2}),
do not yield zero automatically, though annihilate the r.h.s.
So it is a nontrivial consequence of the Poisson structure (49) that
$D_1 \Pi_{\Phi} ({\o D}_1 \Pi_{\Phi^{\dagger}})$ commute with
$\Phi (\Phi^{\dagger})$ at equal supertime, or, in other words, that
the $\theta^1$ (${\o \theta}^1$) dependence of $\Pi_{\Phi}$
($\Pi_{\Phi^{\dagger}}$)
in (\ref{PB2}) can be consistently neglected. This property is
evidently compatible with the superfield equations of motion (43)
(and,
actually, with the equations corresponding to an arbitrary choice of
the chiral superfields potential in (40)). It is
straightforward
to check that on shell the super-Poisson brackets (\ref{PB2})
yield the Poisson brackets obtained from the theory formulated
in
terms of component fields.

As the next crucial step in constructing $N=2$ superhamiltonian
formalism, we define the $N=2$ superhamiltonian by
\a  \label{SH2}
H \;=\; \int dx_Ld\theta^1 {\cal H}_L + \int dx_Rd{{\o \theta}}^1
{\cal H}_R \equiv
H_L+H_R
\b
with
\a
{\cal H}_L \equiv \Pi_{\Phi^\dagger} {\dot \Phi}^{\dagger} -{\cal
L}_L
= -i({\o D}_1\Phi )' \Phi^{\dagger} +{\textstyle{1\over 8}} (
\Pi_{\Phi^{\dagger}}{\o D}_1 \Pi_{\Phi}+ {\o D}_1
\Pi_{\Phi^{\dagger}}
\Pi_{\Phi}) -
{\textstyle {i\over 2}}\Pi_{\Phi}(e^{\Phi^{\dagger}}-
\beta e^{-\Phi^\dagger} )&   \label{HL} \\
{\cal H}_R \equiv
\Pi_{\Phi} {\dot \Phi} -{\cal L}_R =
i ( D_1\Phi^{\dagger} )'\Phi - {\textstyle {1\over 8}} (\Pi_{\Phi}
D_1 \Pi_{\Phi^{\dagger}} + D_1 \Pi_{\Phi}\Pi_{\Phi^{\dagger}})
+{\textstyle {i\over 2}}
\Pi_{\Phi^{\dagger}}
(e^{\Phi}-\beta e^{-\Phi}).& \label{HR}
\b
In the process of obtaining the expressions (\ref{HL}), (\ref{HR}) we
have
eliminated $\dot{\Phi}, {\dot \Phi}^{\dagger}$ with the help of
the kinematical relations
(\ref{Al}).

It is remarkable that $H_L$ and $H_R$ separately satisfy the
following
conservation laws
\a
&& {\o D}_0 H_L = {\o D}_1 H_L= D_0 H_L = {\dot H}_L = 0
\b
and
\a
&& D_0 H_R= D_1 H_R = {\o D}_0 H_R = {\dot H}_R = 0\;.
\b
In both sets first laws are fulfilled kinematically, without
assuming any dynamics for the involved superfields, while the
remaining
ones are genuine conservation laws: they are consequences of the
equations of motion (43). It is an easy exercise to see that they are
compatible
with the algebra of $N=2$ spinor derivatives (\ref{ACD22}).

It is interesting that $H_L$ and $H_R$ can be interpreted as the
time-translation operators, respectively for $\Phi^{\dagger}$ and
$\Phi$:
it is straightforward to check that the pairs of the Hamilton
equations
\a  \label{HE21}
{\dot \Phi}^\dagger &=& 2 \{ H_L, \Phi^\dagger\}\nonumber\\
{\dot \Pi}_{\Phi^\dagger} &=& 2 \{ H_L , \Pi_{\Phi^\dagger} \}
\b
and
\a \label{HE22}
{\dot \Phi} &=& 2 \{H_R , \Phi \}\nonumber\\
{\dot \Pi}_\Phi &=& 2\{ H_R, \Pi_\Phi \}
\b
immediately yield, with respect to the super Poisson
brackets (\ref{PB2}), the identities
(\ref{Al}) and the second equations from the conjugate pairs
(\ref{Con1}),
(\ref{Con2}). In fact, the whole system (\ref{Con1}), (\ref{Con2}) is
reproduced, since all the spinor and ordinary derivatives of the
first
equations vanish as a consequence of the second ones and then the
validity of the first equations is clear by $2D$ Lorentz covariance.

Actually, after passing in $H_R$ to the left- and in $H_L$ to the
right-chiral bases one may evaluate the Poisson brackets of $H_L$
with $\Phi$,
$\Pi_{\Phi}$ and $H_R$ with $\Phi^{\dagger}$, $\Pi_{\Phi^{\dagger}}$
and check
that these brackets are such that the same evolution equations
(\ref{HE21}),
(\ref{HE22}) can be equivalently obtained as those with respect to
the
full superhamiltonian (\ref{SH2})
\a
{\dot \Phi} &=& \{H , \Phi \}\nonumber\\
{\dot \Phi}^\dagger &=& \{ H, \Phi^\dagger\}\nonumber\\
{\dot \Pi}_\Phi &=& \{ H, \Pi_\Phi \}\nonumber\\
{\dot \Pi}_{\Phi^\dagger} &=& \{ H , \Pi_{\Phi^\dagger} \}\;.
\b

The fact that the dynamics is splitted into two conjugate
hamiltonians
leading to two conjugate sets of equations of motion is a peculiar
feature of the second supersymmetry: indeed, $N=2$ supersymmetric
integrable models
can be formulated through two conjugate Lax pairs (see
\cite{IvTop}).

Let us point out that, in contrast with the $N=1$ case, in the
$N=2$ superhamiltonian approach one does not immediately
reproduce the original
Lagrangian
superfield equations of motion (\ref{eqmoN2}), but merely their
consequences
(\ref{Con1}), (\ref{Con2}). The latter system actually is not
fully equivalent
to (\ref{eqmoN2}): these equations are restored from it up to an
arbitrary complex inetgration
constant (of dimension of mass) in their r.h.s., respectively $c$ and
$c^{\dagger}$,
which amounts to adding the linear
terms $-c\Phi$ and $-c^{\dagger}\Phi^{\dagger}$ to the
superpotentials in the
Lagrangians (\ref{S2}), (\ref{SChL}):
$$
L_{L} \Rightarrow L_{L} - c^{\dagger} \Phi^{\dagger}, \;\;
L_{R} \Rightarrow L_{R} - c \Phi\;.
$$
Once this constant is included in the Lagrangian,
it will appear as well in the superhamiltonians (\ref{SH2}) and
(\ref{HL}),
(\ref{HR}):
$$
{\cal H}_{L} \Rightarrow {\cal H}_{L} - {\textstyle {1\over 4i}}
c^{\dagger} \Pi_{\Phi},\;\;
{\cal H}_{R} \Rightarrow {\cal H}_{R} - {\textstyle {1\over 4i}}
c \Pi_{\Phi^{\dagger}}\;.
$$
Just these modified hamiltonians now satisfy the conservation laws
(53), (54): while checking the validity of the latter, one needs to
make use of the original superfield equations of motion, not only
of their consequences (\ref{Con1}), (\ref{Con2}).
Thus we conclude that there exists a family of
$N=2$ superhamiltonians parametrized by a complex parameter $c$,
such
that all
these hamiltonians yield the same dynamical equations (\ref{Con1}),
(\ref{Con2})
and the identities (\ref{Al}) as the associated Hamilton system and
reproduce the original set of equations up to a freedom related to
this parameter. For the
time being the origin and meaning of this ambiguity is not quite
clear for
us. Note that such a deformation of the $N=2$ super Liouville action
has been proposed in
ref. \cite{Porrati} to avoid difficulties with infra-red
divergences. The resulting more general theory was called
the ``$N=2$ Morse - Liouville'' one.

\section{Concluding Remarks}

In this paper we have defined a superhamiltonian formalism for
$N=1,2$ $2D$
theories which naturally extends the standard hamiltonian
formulation.
Its main advantage compared to the standard canonical hamiltonian
treatment of
such theories lies in its manifest supersymmetry: e.g., the full
supersymmetric set of the field equations including those for
auxiliary fields arise as the Hamilton ones.

Our framework allows to introduce Posson structures and perform
computations
with them at a superfield level; therefore we expect it to find
natural applications in the integrable supersymmetric models and in
any
kind of supersymmetric theory where there is a need to
investigate algebraic structures based on Poisson brackets.
{}~\\~\\

\noindent
{\large{\bf Acknowledgements}}
{}~\\~\\
We thank F. Delduc and A. Isaev for useful discussions. E.I. is
grateful to   ENSLAPP, ENS-Lyon, for hospitality.
\setcounter{equation}{0}
{}~\\
{}~\\
{\Large{\bf{{Appendix: Component form of superhamiltonian}}}}
\def\theequation{A.\arabic{equation}}

{}~
\\
{}~\\
In this appendix we present the superhamiltonian
(\ref{SH}) for the $N=1$ sinh-Gordon model in terms of component
fields.

The superfields $\Phi$, $\Pi_\Phi$ have the following component
expansions
\a
\Phi &=& \phi +\theta^0\psi_0 + \theta^1\psi_1 +\theta^0\theta^1
F\nonumber\\
\Pi_\Phi &=& {\textstyle {i\over 2}} (\psi_0 +i\theta^0\phi ' +
\theta^1(F+i{\dot \phi}) +\theta^0\theta^1 (i\psi_1
'-i{\dot{\psi_0}}))
\b
where $\psi_{0,1}$ are fermionic fields, $\phi, F$ bosonic ones ($F$
auxiliary).

The superfield equation of motion (\ref{eqmo}) amounts to the
following set of equations for the component fields
\a
F &=& -i( e^\phi -\beta e^{-\phi})\nonumber\\
{\dot {\psi}}_0 -{\psi_1}' &=& -\psi_0( e^\phi +\beta
e^{-\phi})\nonumber\\
{\dot\psi}_1 -{\psi_0}' &=& \psi_1 (e^\phi +\beta e^{-\phi})
\nonumber\\
{\ddot{\phi}} - {\phi '}'
&=& i F(e^\phi +\beta e^{-\phi}) - i\psi_0\psi_1  (e^\phi -\beta
e^{-\phi}).\label{eqmot}
\b

The hamiltonian density ${{\cal H}}$ can be straightforwardly
computed
from eq.(\ref{SH}). After doing the $\theta^1$ integration there, one
is left
with an integral over the spatial coordinate $x$. Though the
integrand in it bears a dependence on $\theta^0$, the coefficient
before
$\theta^0$ turns out to be a total $x$ derivative and gives
no contribution, in agreement with the first of the conservation laws
(20). So the hamiltonian $H$ is
just given by
\a
H &=& {\textstyle {1\over 4}}\int dx (
(F+i{\dot\phi})^2 -\phi'\phi '
+i\psi_1 '\psi_1 +i \psi_0'\psi_0+
\nonumber\\
&& 2i(F+i{\dot\phi})(e^\phi - \beta e^{-\phi}) -2i\psi_0\psi_1
(e^\phi + \beta e^{-\phi})).
\b
Notice that the fields
$F, {\dot{\phi}}$ enter into this hamiltonian in the combination
$F+i{\dot{\phi}}$. Only after employing the algebraic equation for
the auxiliary field $F$, eq.(A.3) reduces to the standard hamiltonian
constructed directly
in the component fields formalism.

To find out the significance of this deviation from the standard
hamiltonian approach, let us look at the component content of the
super-Poisson brackets (\ref{SPoisson}). The ordinary Poisson
brackets at equal time
implied by (\ref{SPoisson}) for the component fields fall into the
three categories: (i).The standard ones matching with those
constructed in the
component fields canonical formalism; (ii). The relations involving
the auxiliary field $F$; (iii) Some additional
constraints which involve, e.g., $\dot{F}$, $\ddot{\phi}$ and
the validity of which should be checked with making use of the
previous type relations and the equations of motion. For
self-consistency
of the whole set of these Poisson brackets it is essential that
the "super-simultaneity" is defined as in eq. (23).

In the first category we have the following relations (for brevity we
suppress the argument $t$ of fields)
\a
\{ {\dot\phi}(x), \phi (x') \}_- &=& -2\delta (x-x')\nonumber\\
\{ \psi_0 (x) ,\psi_1 (x')\}_+ &=& -2i\delta (x-x')\nonumber\\
\{ \phi(x), \phi(x')\}_- &=& \{ \phi(x), \psi_{0,1}(x') \}_- =
0\nonumber\\
\{ \psi_0(x), \psi_0(x') \}_+& =& \{ \psi_1(x), \psi_1(x') \}_+ = 0
\nonumber\\
\label{compopoisson}
\b
The second category involves the relations
\a
\{ F (x), \phi(x')\}_-= \{ F(x), \psi_{0,1}(x') \}_- &=&0\nonumber\\
\{ F(x) + i\dot{\phi}(x), F(x') +i\dot{\phi}(x') \}_- &=& 0.
\b
An example of the constraints of the third category is supplied by
\a
\{ {\dot\psi_0}(x), \psi_0(x')\}_+ &=& 2i \partial_x \delta (x-x').
\b

Now it is easy to see that the whole set of the component equations
(A.2)
amounts to the set of the Hamilton equations
\a
\dot{B} = \{H, B\} \nonumber
\b
with, respectively, $B = \phi, \psi_0, \psi_1, F+i\dot{\phi}$. So
only
the first and second category of the component Poisson brackets are
really
of need to derive the equations of motion{\footnote{Actually,
for this derivation we merely need to know the Poisson brackets just
between
the fields $\phi, \psi_0, \psi_1, F+i\dot{\phi}$, but not, e.g.,
between
$F$ and $F$, $F$ and $\phi$ separately.}. The remarkable difference
from the
standard component hamiltonian formalism lies in the fact that the
{\it whole} set of field equations including the equation for the
auxiliary field $F$ follows as the
Hamilton equations with respect to the hamiltonian (A.3). This
comes about just because (A.3) differs from the standard component
hamiltonian by terms which vanish upon exploiting the equation for
$F$.

All these features are independent of the specific form of the chosen
superpotential and
therefore are valid for a more general class of theories than those
explicitly treated in the present paper.

\end{document}